\documentclass[prl,twocolumn,showpacs,amsmath,amssymb,superscriptaddress]
{revtex4}

\usepackage[dvips]{color}
\usepackage{graphicx}

\begin{document}

\title{Disorder-enhanced phase coherence in \\
       trapped bosons on optical lattices}
  
\author{Pinaki Sengupta}
\affiliation{Department of
    Physics \& Astronomy, University of Southern California, Los Angeles,
    California, 90089, USA}
\author{Aditya Raghavan}
\affiliation{Department of
    Physics \& Astronomy, University of Southern California, Los Angeles,
    California, 90089, USA}
\author{Stephan Haas}
\affiliation{Department of
    Physics \& Astronomy, University of Southern California, Los Angeles,
    California, 90089, USA}

\begin{abstract}
The consequences of 
disorder on interacting bosons trapped in
optical lattices are investigated by 
quantum Monte Carlo simulations. 
At small to moderate strengths of potential 
disorder a unique effect is observed: if there is a 
Mott plateau at the center of the trap in the clean limit, 
phase coherence {\it increases} as a result of
disorder. The localization effects due to correlation and disorder
compete against each other, resulting in a partial delocalization
of the particles in the Mott region, which in turn leads to increased 
phase coherence. 
In the absence of a Mott plateau, this effect is absent.
A detailed analysis of the uniform system without a trap shows that the 
disordered states participate in a Bose glass phase.
\end{abstract}


\maketitle

Recent advances in experiments with ultracold atoms in magneto-optical
traps have opened a new frontier in the study of strongly correlated
systems. Some of the more remarkable early
experimental breakthroughs include the realization of Bose-Einstein
condensation\cite{BEC-1995} and fermionic superfluidity.\cite{Fermi-SF-1999}
More recent achievements include tuning a degenerate Fermi gas across BCS-BEC
crossover via Feshbach resonance,\cite{BCS-BEC} and detecting 
superfluid-to-Mott insulator transitions of trapped bosons in optical 
lattices.\cite{Greiner-2002,Stoferle-2004} 
An unprecedented control over experimental parameters in these systems makes
them ideally suited for studying many-body phenomena.

Until recently, the majority of experiments with trapped ultracold atoms
have been performed on clean systems. 
Indeed, the ability to create perfect (defect free) optical lattices  
is a major advantage over condensed matter experiments. On the other hand,
the complete control over the trapping potentials makes it possible to
introduce disorder and tune the disorder strength in a controlled
fashion. The exciting possibilities of being able to study
novel disorder-related phenomena such as Anderson 
localization and to explore novel quantum glassy phases have made the 
investigation of disorder in these systems an area of emerging
interest.

Disorder can be generated in optical lattices by exposure to speckle 
lasers\cite{Horak-2000,Clement-2005}, adding an incommensurate 
lattice-forming 
laser\cite{Damski-2003,Sanpera-2004}, or other means\cite{Folman-2002}.
It is thus possible to investigate different disorder-induced phenomena in
a controllable manner, in contrast e.g. to previous studies with granular
superconductors or $^4$He in vycor glass. The interplay between
disorder and interactions in trapped Bose-Einstein condensates  
has recently been explored experimentally in $^{87}$Rb, both in the
continuum\cite{Lye-2005,Fort-2005,Clement-2005} and in an 
optical lattice\cite{Schulte-2005}. On the theoretical front,
such systems have been investigated within the frameworks of mean-field 
theories,\cite{Damski-2003,Schulte-2005,Clement-2005,Kuhn-2005}
Bose-Fermi mapping,\cite{DeMartino-2005} and the transfer
matrix formalism.\cite{Gavish-2005}

\begin{figure}
\includegraphics[width=6.0cm]{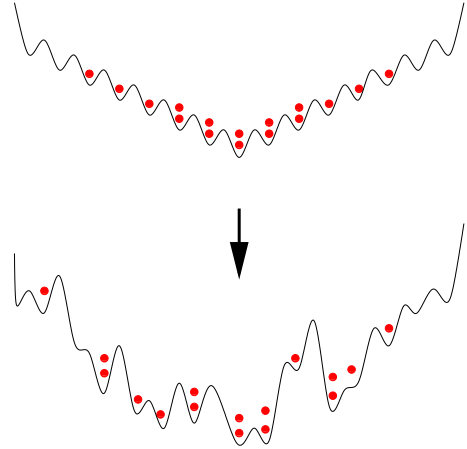}
\vskip1mm
\caption{Illustration of atoms in a potential trap. In the absence of 
interactions between the atoms, disorder leads to an immediate 
localization of the single-particle wave 
functions.
}
\label{fig:fig1}
\end{figure}

In this paper, we study interacting trapped bosons in one-dimensional 
optical lattices with a tunable random potential using a quantum
Monte Carlo method. Previous investigations have argued that under 
realistic conditions, such diagonal disorder dominates over randomness 
in the hopping amplitude.\cite{Damski-2003}  In experiments with optical 
lattices, the primary source of information about the state of the system 
arises from the analysis of the momentum distribution function, obtained 
from matter-wave interference after the release of the trap and subsequent 
free evolution of the particles. In view of this, here we focus on the 
signatures of disorder on the momentum distribution. The simulations 
presented here predict a unique enhancement of phase coherence at small to
moderate strengths of disorder, when the groundstate has a Mott 
insulating region, i.e. a Mott plateau, at the center of the trap. 
Since Mott insulating states are now experimentally observable, and 
disorder strengths are controllable by tuning the intensity of the speckle 
laser, these results can be straightforwardly tested using currently available  
experimental methods. One technical difficulty in using current setups is
the typically long wavelength of speckle patterns (typically 8-10 times the optical
lattice spacing) and the limited size of optical lattices (40-65 lattice
spacings along each axis). However, these difficulties are likely to be 
overcome in the near future with advances in realizing larger lattices
and / or using more than one speckle laser to create disordered potentials
with a shorter correlation.

Bosons in an optical lattice are well described by the one-band Bose Hubbard
model.\cite{Jaksch-1998} In the presence of a
strong periodic (lattice) potential, single-particle Wannier functions
localized on the lattice sites form a complete basis set. 
Interactions are typically not strong enough to excite higher
vibrational bands, justifying the single band approximation. Hence the
low-energy physics of trapped bosons in a one-dimensional  optical 
lattice can be described by the Hamiltonian
\begin{eqnarray}
H &=& -t\sum_{i=1}^L(b_{i+1}^{\dagger}b_i + h.c.)+ 
{U\over 2}\sum_{i=1}^L n_i(n_i - 1)  - \mu_0\sum_{i=1}^Ln_i\nonumber \\
 & & + V_Ta^2\sum_{i=1}^L(i-{L\over 2})^2n_i + \sum_{i=1}^LW_in_i.
\label{eq:bhm}
\end{eqnarray}  
Here $b_i^{\dagger}(b_i)$ creates (annihilates) a boson at site $i$, 
$n_i=b_i^{\dagger}b_i$ is the number operator, and $U$ is the strength 
of the repulsive on-site Hubbard interaction between bosons.
The bare chemical potential $\mu_0$ controls the filling of the lattice, 
$V_T$ is the strength of the trapping potential, $a$ denotes the lattice 
spacing, and $W_i$ introduces diagonal disorder in the form of a random 
site energy. 
The hopping amplitude $t$ is set equal to unity in the simulations, and 
all other parameters are implicitly
expressed in units of $t$. In the experiments, $W_i$ is
commonly realized by a speckle laser. 
Furthermore, recent measurements
indicate significantly correlated disorder, with a correlation length that 
is typically several times longer than the optical lattice spacing,
due to the diffraction-limited imaging
of the speckles coming from a diffusion plate onto the 
trapped atoms.\cite{Lye-2005,Fort-2005}
In view of this,
we consider finitely correlated disorder with a 
random potential $W_i$ that is
distributed as 
\begin{equation}
\overline{W_l} = 0,\hskip0.5in \overline{W_lW_{l'}} = \Delta \delta_{l,l'}.
\end{equation}
The overbar denotes averaging over disorder realizations, and $\Delta$
is disorder strength. The indices $l$ are measured in units of $5a$.

\begin{figure}
\includegraphics[width=8.3cm]{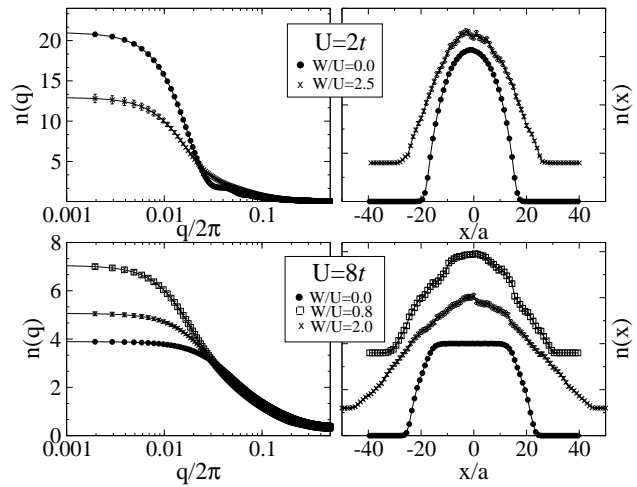}
\vskip1mm
\caption{Effects of disorder on the momentum distribution $n(q)$ and 
on the real-space density profile $n(x)$
in the weak (upper panels) and strong (lower panels) coupling limit. 
The density 
profiles are vertically offset to enhance clarity of presentation. 
} 
\label{fig:nqnx}
\end{figure}

In the absence of disorder, the Bose Hubbard model (\ref{eq:bhm}) in a 
trap has been extensively studied.\cite{BHM} Due to the spatially varying 
trapping potential, the system is never in a uniform phase. Depending on 
the strength of the trapping potential, $V_T$, the on-site interaction, 
$U$, and the density of particles, $n$, the groundstate   
can belong to one of two classes. At sufficiently small $U/t$ and total
density, $n$, it consists of a superfluid (SF) domain extending across 
the entire system. The site-occupation, $n_i$
varies continuously from zero at the edges to a finite maximum value 
at the center of the trap. The momentum distribution, $n(q)$, features
a sharp peak at $q=0$ reflecting strong phase coherence across the
entire system. The full width at half maximum of the zero-momentum
peak is inversely proportional to the correlation length $\xi$. In a 
thermodynamic SF $\xi$ diverges, whereas in the one-dimensional confined 
SF $\xi\propto N$ and $n(q=0)\propto \sqrt{N}$. 

On the other hand, at larger $U/t$ and sufficiently
large $n$, the groundstate contains one or more Mott insulating (MI)
regions with integer $n_i$, along with domains of SF at the edges of the 
trap and in between the MI regions (if there are more than one such 
regions).\cite{BHM} In the present study, we consider the simplest case 
of one MI plateau at the center of the trap and two SF domains at the 
edges -- the local density profile consists of an extended region at the 
center with $n_i=1$  and two regions with $0<n_i<1$ near the trap edges.
In the presence of MI region(s), the range of coherence is greatly reduced, 
and the momentum distribution shows a weak peak at $q=0$, and a subsequent
slow decrease in $n(q)$ as a function of $q$. Depending on whether there
exists an MI region at the center of the trap, we find the effects of disorder
can be markedly different. 

We use the stochastic series expansion quantum Monte Carlo 
method to simulate the disordered Bose-Hubbard model on finite-sized 
lattices.\cite{Sandvik-1999} The density profiles, $n(x)$, and the
momentum disribution function, $n(q)$, are measured within the grand 
canonical ensemble, using sufficiently large values of the inverse 
temperature, $\beta$, in order to obtain groundstate properties. 
Disorder averages over 400-2000 realizations are performed, depending 
on the parameters. The lattice sizes are chosen to be sufficently
large to ensure that the local density vanishes at the edge of the trap.
We found that a lattice size $L=120$ is necessary for large values
of $U(\ge 8t)$ at strong disorder ($W/U > 1.5$), whereas $L=80$ was
sufficient for the other parameter sets. Results are displayed for 40 
bosons in a parabolic trap with $V_Ta^2=0.015t$ for a range of values
of $U/t$, ranging from the weak to strong coupling limit.

Fig.~\ref{fig:nqnx} shows the consequences of a disordered site-potential
on the momentum distribution function $n(q)$ and the real-space density 
profile $n(x)$ for two values of the on-site interaction, $U/t$. The
upper panels show the results for $U/t=2$, when the groundstate in the 
clean limit ($W=0$) consists of a single SF domain extending acroos the 
trap. Adding disorder produces localization centers around which the
wave functions get localized. This in turn reduces coherence and hence 
results in a smaller zero-momentum peak. At sufficiently weak disorder 
strengths, these effects are found to be relatively small in finite 
systems, and the groundstate retains a predominantly SF character 
-- the deviation of the momentum distribution from its clean limit 
is exponentially small. However, with increasing disorder strength, 
the localization effects strengthen, and hence the momentum distribution 
starts to deviate significantly, i.e. the peak at $n(q\rightarrow 0)$
becomes shorter and broadened. 

On the other hand, for $U/t=8$ (lower panels), the groundstate of 
the system, in the clean limit, has a Mott plateau at the center 
and the effects of disorder are considerably more interesting. The 
wave functions in the MI region in the absence of disorder are 
localized due to the strong Hubbard interactions. The onset of 
disorder weakens the correlation between particles, and thus leads
to a partial delocalization of the strongly localized Wannier 
orbitals at small to moderate strengths of disorder. Phase coherence 
hence {\it increases} as the MI region ``melts'', reflected in a 
stronger $q=0$ peak in $n(q)$ with increasing disorder. For larger 
disorder potentials, the Anderson localization effects dominate, 
and the zero-momentum peak starts to decrease.

\begin{figure}
\includegraphics[width=8.3cm]{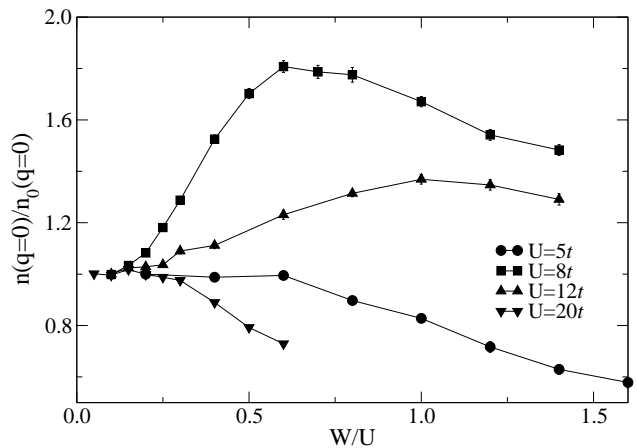}
\vskip1mm
\caption{Dependence of the zero-momentum peak height, $n(q=0)$, on the disorder
strength $W$
for different values of $U/t$. At $U/t=5$, the groundstate consists of a single
SF domain, whereas for the other values, there exist a finite MI region at the 
center of the trap.} 
\label{fig:n0}
\end{figure}

Fig.\ref{fig:n0} shows the variation of the strength of the zero-momentum 
peak as a function of disorder for four different values of $U/t$, ranging
from the weak to the strong coupling limit. For intermediate values 
of the interaction strength,
the peak at $n(q=0)$ increases at small to intermediate disorder (reflecting 
the melting of the central MI region), reaches a maximum, and finally decreases
continuously for larger $W$. This is to be contrasted with 
the weak coupling limit, 
where $n(q=0)$ remains fairly unchanged for small $W$, and then
decreases rapidly 
at larger disorder strengths. 
The exact position of the maximum and its value, $n_{max}(q=0)$
depends on 
$U/t$, $V_T/t$, and $n$. The
peak position shifts to larger disorder strengths with increasing Hubbard
intearction strength,
indicating a competition between Mott localization and Anderson localization
that depends on the ratio of $W$ vs. $U$.
The 
maximum height, $n_{max}(q=0)$ and its ratio to the clean system value, 
$n_0(q=0)$,
decrease rapidly with $U/t$. At $U/t=20$, i.e. close to the hard-core
limit, no visible increase with $W$ is noticable in $n(q=0)$. 

\begin{figure}
\includegraphics[width=8.3cm]{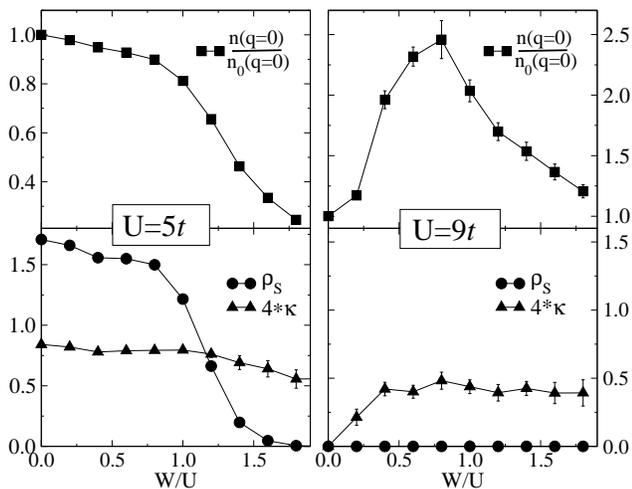}
\vskip1mm
\caption{Dependence of the momentum distribution function at
$q = 0$, the stiffness and the compressibility
of SF (left panels) and MI (right panels) phases on disorder strength. For the
SF phase, the groundstate is predominantly SF at small to moderate disorder and a
Bose glass at large disorder. For the MI phase, the groundstate is a Bose glass
at all strengths of disorder studied. The strength of the 
zero-momentum peak varies
non-monotonically, but the groundstate always remains a Bose glass.} 
\label{fig:uni}
\end{figure}

What is the nature of the state that results from the onset of disorder?
Does the partial delocalization of the MI states result in a SF or a glassy phase? 
To answer these questions, we analyze the effects of disorder in the absence of a 
trapping potential. At $V_T=0$, the groundstate is in a pure phase, and 
can be conveniently characterized by measuring {\it global} observables, 
such as the compressibility and the stiffness. 
For shallow trapping potentials, the system is adiabatically connected to the
periodic case, such that the domains of the different phases
retain their global characteristics to a large extent. This allows us
to study the effects of disorder in the pure phases and extend the results to
corresponding domains in the trapped system. 
Let us focus on the global compressibility, 
$\kappa=\beta(\langle n^2\rangle - \langle n\rangle^2$,
and the stiffness, $\rho_s$, to
differentiate between various possible phases -- SF, MI or Bose glass (BG).
The stiffness gives the response of the system to a uniform twist in the
spatial boundary as $\rho_s=\partial^2 E/\partial \phi^2$, under which
the kinetic energy term in the Hamiltonian (\ref{eq:bhm}) is replaced
by $-t\sum (e^{-i\phi}b_{i+1}^\dagger b_i + h.c.)$. In practice, the
stiffness is simply related to the fluctuations in the winding number in
the simulations as $\rho_s=\langle W^2\rangle/2\beta L$.\cite{Pollock-1987}
The results shown in Fig.~\ref{fig:uni} are for uniform SF (left panels) 
and MI (right panels) phases in the presence of
potential disorder. The dependence of $n(q=0)$ on $W$ is
qualitatively similar compared with the case of trapped atoms. 
For the SF phase, $n(q=0)$ remains 
practically unchanged at small $W$, and starts to decrease monotonically for
larger disorder. On the other hand, in the MI phase
$n(q=0)$ first increases and then 
decreases with increasing $W$, confirming that the 
disordered states are qualitatively similar to their counterparts in the 
presence of a trapping potential.
The stiffness and compressibility data allow us to further characterize
the disordered states. For the SF groundstate, both $\rho_s$ and $\kappa$
are finite for the range of $W$ over which $n(q=0)$ remains close to its
$W=0$ value, consistent with a predominantly SF character of the disordered 
state. At larger $W$, as $n(q=0)$ starts to deviate significantly, 
the compressibility
remains finite, but the stiffness rapidly decreases to zero. Hence the 
groundstate at large disorder is a compressible insulator, 
i.e. a {\it Bose glass}
(BG).
From the present data 
it is not clear if one needs a finite critical disorder to destroy superfluidity
or simply the localization length at small disorder is larger than the system
sizes considered here. For the MI phase, both $\rho_s$ and $\kappa$ vanish 
in the clean system. With the onset of disorder, the stiffness remains
zero, but the compressibility acquires a  finite value. The disordered MI is 
thus a BG at all strengths of disordered considered here.  

In summary, based on quantum Monte Carlo simulations
ensembles of interacting trapped atoms are found to
respond in a highly
non-trivial fashion to diagonal disorder. If the clean system consists
of a single superfluid domain, the particles are simply localized by 
the random potential. This is reflected by progressive suppression of
the $q\rightarrow 0$ peak in the momentum distribution function. On the 
other hand, if the clean system contains Mott insulating regions the  
response to on-site disorder is non-monotonic, and depends on the relative
strength of the Hubbard interaction and the disorder potential. Because
of this competition there is an intermediate regime with enhanced phase
coherence. 
An analysis of the compressibility and stiffness in the corresponding 
periodic systems, i.e. in the 
absence of a trapping potential, reveals that the resulting disordered state 
in either case is a Bose glass. The predicted enhancement of phase coherence 
by diagonal disorder suggests an interesting extension of recent experiments
on optical traps in the presence of a tunable speckle laser.\cite{Fort-2005}

\noindent
\underbar{Acknowledgments} We are grateful to M.~Olshanii, O.~Nohadani and 
T.~Roscilde for fruitful 
discussions, and acknowledge support by the DOE under grant DE-FG02-05ER46240. 
The simulations were carried out at the HPC Center at USC.

\end{document}